\documentclass{PoS_preprintnumbers}

\usepackage{amscd,amsfonts}
\usepackage{amsmath,amssymb}
\usepackage{amsxtra,mathrsfs}
\usepackage{multicol,multirow}
\usepackage{axodraw4j}
\usepackage{shuffle}

\usepackage{epstopdf}

\title{Color decomposition of multi-quark one-loop QCD amplitudes}

\ShortTitle{Color decomposition of multi-quark one-loop QCD amplitudes}

\author{\speaker{Christian Reuschle}\\
        Institute for Theoretical Physics, Karlsruhe Institute of Technology, 76128 Karlsruhe, Germany.\\
        E-mail: \email{christian.reuschle@kit.edu}
}
\author{Stefan Weinzierl\\
        PRISMA Cluster of Excellence,\\ Institute of Physics, Johannes Gutenberg University Mainz, 55099 Mainz, Germany.\\
        E-mail: \email{stefanw@thep.physik.uni-mainz.de}
}


\Preprint{KA-TP-16-2014, MITP/14-043}

\abstract{
In this talk we discuss the color decomposition of tree--level and one--loop QCD amplitudes with arbitrary numbers of quarks and gluons. We present a method for the decomposition of partial amplitudes into primitive amplitudes, which is based on shuffle relations and is purely combinatorial \cite{Reuschle:2013qna}. Closed formulae are derived, which do not require the inversion of a system of linear equations.
}

\FullConference{Loops and Legs in Quantum Field Theory\\
		27 April 2014 - 02 May 2014\\
		Weimar, Germany}

\begin{document}

\newcommand{\ibar}{\bar{\imath}}
\newcommand{\jbar}{\bar{\jmath}}
\newcommand{\qbar}{\bar{q}}
\newcommand{\reference}{\textcolor{red}{[...]}}
\newcommand{\scriptA}{\mathcal{A}}
\newcommand{\si}{\sigma}

\section{Introduction}

Due to the quite complex nature of this topic we limit ourselves to a rather basic description of our method, trying to grant a more intuitive insight. For a detailed treatment we refer the reader to an elaborate discussion of our method in \cite{Reuschle:2013qna}.

Color decomposition offers a systematic tool to deal with the most complicated color structures in QCD amplitudes with large numbers of particles. 
In the so--called color--flow decomposition \cite{colorflow} the factorization of color information and kinematical information,
\begin{align}
\scriptA = \sum_i C_iA_i\;,
\end{align}
is thereby such that the color coefficients $C_i$ are products of open and closed color strings, given by
\begin{align}
c_{\mathrm{closed}}(g_1,g_2,...,g_n) &= \delta_{i_{g_1}\jbar_{g_2}}\delta_{i_{g_2}\jbar_{g_3}}...\delta_{i_{g_n}\jbar_{g_1}}\;, \nonumber\\
c_{\mathrm{open}}(q,g_1,g_2,...,g_n,\qbar) &= \delta_{i_q\jbar_{g_1}}\delta_{i_{g_1}\jbar_{g_2}}\delta_{i_{g_2}\jbar_{g_3}}...\delta_{i_{g_n}\jbar_{\bar{q}}}\;,
\end{align}
where the empty closed string is given by $c_{\mathrm{closed}}()=N$.
The partial amplitudes $A_i$ are gauge invariant subsets of color--stripped diagrams, which are by definition color--ordered and are built from color--stripped Feynman rules (to be found for example in \cite{Reuschle:2013qna,Weinzierl:2005dd}). The color--stripped Feynman rules are by construction antisymmetric in the exchange of the external legs, which means that also the diagrams in the partial amplitudes are distinguished by the cyclic orderings of the external legs around the diagrams (usually defined in clockwise direction).
One peculiarity of the color--flow decomposition is the decomposition of the gluon field in terms of its natural matrix representation, which leads to the decomposition of the $SU(N)$ gluon propagator in terms of a $U(N)$ part and a $U(1)$ part and is schematically depicted by
\begin{align}
\raisebox{-0.4cm}{\includegraphics[scale=0.75]{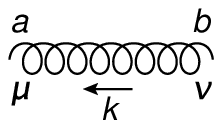}} 
\quad\hat{=}\quad 
\raisebox{-0.4cm}{\includegraphics[scale=1.4]{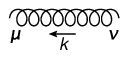}} 
\;\;\times\;\; 
\left(\;\; 
\raisebox{-0.35cm}{\includegraphics[scale=1.0]{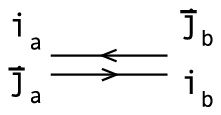}} 
\;\;-\;\frac{1}{N}\;\; 
\raisebox{-0.35cm}{\includegraphics[scale=1.0]{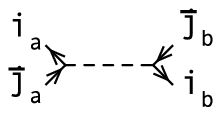}} 
\;\;\right) \;,
\end{align}
which is well known from the $1/N$ expansion of an $SU(N)$ theory. The kinematical parts of the $U(N)$ and $U(1)$ gluon propagators are thereby identical, whereas the color parts are different. Since the color coefficients in the color--flow decomposition are simple products of Kronecker deltas of fundamental and antifundamental color indices, the contraction of the color indices, upon squaring the amplitude, becomes trivial. The color--flow decomposition allows for a quite natural picture in terms of all possible flows of color through an amplitude and for an efficient organization in the number of $U(1)$ gluons.

Partial amplitudes are in general not cyclic ordered, which means that not all diagrams in a partial amplitude possess the same cyclic ordering of the external legs: We decompose the partial amplitudes $A_i$ further into so--called primitive amplitudes $P_j$ through
\begin{align}
\scriptA = \sum_i C_iA_i = \sum_i C_i \sum_j F_{ij}P_j \;.
\end{align}
The primitive amplitudes $P_j$ are gauge invariant subsets of color--stripped diagrams, which are by definition cyclic ordered and are built from color--stripped Feynman rules. 
At the one--loop level, primitive amplitudes are further characterized by the particle content in the loop (either closed fermion loops or loops with at least one gluon or ghost particle), and by the position of the loop with respect to the external fermion lines upon following their fermion--flow arrow (a fermion line can either be right-- or left--routing). 
Due to their cyclic ordering, primitive amplitudes can be efficiently constructed by means of recursion relations, 
and they contain in general smaller numbers of kinematical invariants. Many modern methods to compute the virtual contribution in a next--to--leading order computation, such as the virtual subtraction method \cite{virtualsubtraction} or unitarity based techniques \cite{unitaritymethod}, utilize cyclic ordered primitive amplitudes.
To find the linear relations $F_{ij}$ between the $A_i$ and the $P_j$ (i.e. to find all cyclic orderings to a specific color flow) in the general case of QCD amplitudes with arbitrary numbers of quarks and gluons is non--trivial: In the case of tree--level amplitudes, the cyclic ordering is destroyed by the presence of tree--level like $U(1)$ gluons. In the case of one--loop amplitudes the cyclic ordering is destroyed by the additional presence of $U(1)$ gluons in the loop, or because of an additional closed string.

For the cases of tree--level amplitudes with only gluons or with one quark pair plus gluons, the decomposition into primitive amplitudes is trivial: In these cases the partial amplitudes are also primitive. 
Also at the one--loop level closed formulae exist for the cases of amplitudes with only gluons or with one quark pair plus gluons \cite{oneloopcolordecompsimple}.
For the cases of one--loop QCD amplitudes with arbitrary numbers of quarks and gluons approaches exist, which are based on Feynman diagrams and the inversion of systems of linear equations \cite{oneloopmultiquarkpragmatic}. However, from a certain point of view these approaches are somewhat unsatisfactory, since one has to solve (large) systems of linear equations 
and because they rely on Feynman diagrams, which is unaesthetic, as all other parts of a next--to--leading order calculation can be performed without resorting to Feynman diagrams.

The method which we present here avoids Feynman diagrams and the inversion of a system of linear equations, 
but utilizes (generalized) shuffle relations to derive results for the decomposition of one--loop amplitudes with an arbitary number of quarks and gluons in closed form. 

At this point we would like to mention a few other quite recent efforts, which are related to our work: 
Closed results for the decomposition of multi--quark one--loop amplitudes were recently derived in \cite{Schuster:2013aya} as well. 
Relations between multi--quark tree--level primitive amplitudes, based on Dyck words, were recently derived in \cite{Melia:2013bta}. 
General symmetry properties of color structures in one--loop amplitudes, where all fields are in the adjoint representation, were recently studied in \cite{Kol:2014sla}.

\section{Basic tree-level operations}

In the following we will elaborate on some necessary basic operations. Consider therefor the alphabet $A=\{l_i\}=\{q_1,q_2,...,\qbar_1,\qbar_2,...,g_1,g_2,...\}$, made of quark, antiquark and gluon indices: A word $w=l_1...l_k$ is an ordered sequence of letters $l_i\in A$. Consider further ordered sequences up to cyclic permutations: $l_1l_2...l_k \sim l_2...l_kl_1$. A cyclic word $w=[l_1...l_k]$ is then the corresponding equivalence class, denoted by a square bracket around a representative sequence.
We define the shuffle product between two words $w_1=l_1...l_k$ and $w_2=l_{k+1}...l_r$ by
\begin{align}
w_1 \shuffle w_2 = \sum\limits_{\mathrm{shuffles}\;\sigma}l_{\sigma(1)}...l_{\sigma(r)} \;,
\end{align}
where the sum runs over all permutations, which preserve the relative order of $l_1...l_k$ and $l_{k+1}...l_r$, while permutations, where the crossing of fermion lines cannot be avoided, are excluded. Further we define the shuffle product between two cyclic words $w_1=[l_1...l_k]$ and $w_2=[l_{k+1}...l_r]$ by
\begin{align}
w_1 \circledcirc w_2 = \sum\limits_{(\mathrm{cyclic\,shuffles}\;\sigma)/Z_r}[l_{\sigma(1)}...l_{\sigma(r)}] \;,
\end{align}
where the sum runs over all permutations, which preserve the relative cyclic order of $l_1...l_k$ and $l_{k+1}...l_r$, while permutations, where the crossing of fermion lines cannot be avoided, as well as cyclic permutations of $l_{\sigma(1)}...l_{\sigma(r)}$ are excluded. 
We then define pimitive amplitudes $P(w)$ as linear operators on the vector space of cyclic words $w$:
\begin{align}
\sum\limits_{w\in \lambda_1w_1+\lambda_2w_2}P(w)=\lambda_1P(w_1)+\lambda_2P(w_2) \;.
\end{align}
Due to the antisymmetry of the color--stripped vertices the $P(w)$ have a reflection identity $P(w)=(-1)^nP(w^T)$, where $w^T:w=[l_1...l_n] \mapsto w^T=[l_n...l_1]$, and obey partial reflection as well, e.g. $P(q_1\qbar_1q_2\qbar_2)=-P(q_1\qbar_1\qbar_2q_2)$. 

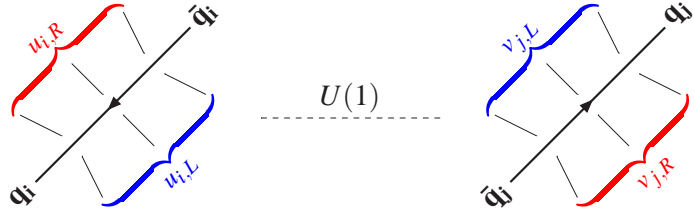
\begin{figure}
\vspace{-1.0cm}
\centering
\rotatebox{45}{\scalebox{0.55}{
\begin{picture}(200,100)(0,0)
\SetArrowInset{0.0}
\SetWidth{1.5}
\ArrowLine(175,50)(25,50)
\SetWidth{0.5}
\Line(60,60)(50,90)
\Line(100,60)(100,90)
\Line(140,60)(150,90)
\Line(60,40)(50,10)
\Line(100,40)(100,10)
\Line(140,40)(150,10)
\Text(20,50)[r]{\huge$\mathbf{q_i}$}
\Text(180,50)[l]{\huge$\mathbf{\qbar_i}$}
\Text(100,95)[b]{\Huge$\textcolor{red}{\overbrace{\textcolor{white}{\hspace{3.75cm}}}^{u_{i,R}}}$}
\Text(100,5)[t]{\Huge$\textcolor{blue}{\underbrace{\textcolor{white}{\hspace{3.75cm}}}_{u_{i,L}}}$}
\end{picture}
}}
\hspace{-0.75cm}
\scalebox{0.45}{\raisebox{2.5cm}{
\begin{picture}(200,100)(0,0)
\DashLine(25,50)(175,50){5}
\Text(100,80)[t]{\Huge $U(1)$}
\end{picture}
}}
\hspace{-0.75cm}
\rotatebox{45}{\scalebox{0.55}{
\begin{picture}(200,100)(0,0)
\SetArrowInset{0.0}
\SetWidth{1.5}
\ArrowLine(25,50)(175,50)
\SetWidth{0.5}
\Line(60,60)(50,90)
\Line(100,60)(100,90)
\Line(140,60)(150,90)
\Line(60,40)(50,10)
\Line(100,40)(100,10)
\Line(140,40)(150,10)
\Text(20,50)[r]{\huge$\mathbf{\qbar_j}$}
\Text(180,50)[l]{\huge$\mathbf{q_j}$}
\Text(100,95)[b]{\Huge$\textcolor{blue}{\overbrace{\textcolor{white}{\hspace{3.75cm}}}^{v_{j,L}}}$}
\Text(100,5)[t]{\Huge$\textcolor{red}{\underbrace{\textcolor{white}{\hspace{3.75cm}}}_{v_{j,R}}}$}
\end{picture}
}}
\vspace{-0.75cm}
\caption{\footnotesize $U(1)$--like connection between the quark lines $i$ and $j$ of two cyclic words $u$ and $v$, respectively, where $u_{i,R}$, $u_{i,L}$, $v_{j,R}$ and $v_{j,L}$ may contain further quark and antiquark indices, as well as gluon indices.}
\label{fig:U1connection}
\end{figure}

For our purposes we need to generalize the above, in order to connect two cyclic words $u$ and $v$ by a $U(1)$ gluon. Consider first that we want to connect a quark line $i$ in $u=[q_iu_{i,R}\qbar_iu_{i,L}]$ with a quark line $j$ in $v=[q_jv_{j,R}\qbar_jv_{j,L}]$, as depicted in fig.~\ref{fig:U1connection}, where $u_{i,R}$ and $u_{i,L}$ ($v_{j,R}$ and $v_{j,L}$) denote the remaining sequences to the right and left of $q_i$ in $u$ ($q_j$ in $v$) in clockwise direction. We can then define a generalized shuffle product by 
\begin{align}
U_{ij}(u,v)=\sum\limits_{\stackrel{(\mathrm{cyclic\,shuffles}\;\sigma)/Z_r}{(q_i...\bar{q_i}...q_j...\bar{q_j}...)}}[l_{\sigma(1)}...l_{\sigma(r)}] \;,
\end{align}
where the sum runs over all permutations, which preserve the relative cyclic order of $l_1...l_k$ and $l_{k+1}...l_r$, while permutations, where the crossing of fermion lines cannot be avoided, as well as cyclic permutations of $l_{\sigma(1)}...l_{\sigma(r)}$ and permutations with cyclic order $\neq[q_i...\bar{q_i}...q_j...\bar{q_j}...]$ are excluded. It can be shown that this corresponds indeed to a $U(1)$--like connection between the quark lines $i$ and $j$ of two cyclic words $u$ and $v$ respectively \cite{Reuschle:2013qna}. The definition of $U_{ij}(u,v)$ is not unique, reflecting the fact that the decomposition of partial amplitudes into primitive amplitudes is not unique. Now suppose that $u$ contains $n_{q_u}$ quark pairs labeled by $ i \in K=\{1,2,...,n_{q_u}\}$ and that $v$ contains $n_{q_v}$ quark pairs labeled by $j \in L=\{n_{q_u}+1,n_{q_u}+2,...,n_{q_u}+n_{q_v}\}$.
We then set
\begin{align}
U(u,v)=\sum\limits_{i\in K}\sum\limits_{j\in L}U_{ij}(u,v) \;.
\label{def_U1_shuffle}
\end{align}
The definition of $U(u_1,...,u_r)$ with more than two arguments is more involved. Suppose that $u_i$ contains $n_{q_i}$ quark pairs labeled from $n_{q_1}+...+n_{q_{i-1}}+1$ to $n_{q_1}+...+n_{q_{i-1}}+n_{q_i}$. 
The operation $U(u_1,...,u_r)$ is then defined by the following algorithm:
\vspace{-1.0ex}
\begin{enumerate}
\itemsep-0.5ex
\item Draw all connected tree diagrams with $r$ vertices, labeled by $u_1$, ..., $u_r$.
\item For every edge connecting $u_k$ and $u_l$ use an operation $U_{ij}$, where $i$ labels a quark pair in $u_k$ and $j$ labels a quark pair in $u_l$, and sum over $i$ and $j$. The order in which these operations are performed is irrelevant, since the operations for a given graph are associative.
\item Sum over all diagrams.
\end{enumerate}
The reason for this definition is that a naive iteration of the operation with two factors in eq.~(\ref{def_U1_shuffle}) is not associative and the corresponding result would not only reduce to diagrams that connect all quark lines in all possible ways by two $U(1)$ gluons, but also take unwanted diagrams into account.

Another possibility to define the operation $U_{ij}(u,v)$ is the following: 
Imagine an ordered amplitude, corresponding to a cyclic word $u=[q_iu_{i,R}\qbar_iu_{i,L}]$ as depicted by the left diagram in fig.~\ref{fig:U1connection}, with two sets $u_{i,R}$ and $u_{i,L}$ of color--connected particles, separated by a parton line $i$. 
Suppose we want to have both sets on one side of the parton line $i$, such that the members of one set never directly couple to the members of the other set: Using shuffle relations and the antisymmetry of the color--stripped vertices we can ``flip'' the elements of $u_{i,L}$ to the other side of the parton line $i$, such that diagrams, where the elements of $u_{i,R}$ and $u_{i,L}$ would couple, 
cancel each other. In terms of primitive amplitudes, this is known as Kleiss--Kuijf relations:
\begin{align}
P_n^{(0)}(u)=(-1)^{n_{u_{i,L}}}\sum\limits_{w\in[q_i(u_{i,R} \shuffle u_{i,L}^T)\qbar_i]}P_n^{(0)}(w) \;,
\end{align}
where $n_{u_{i,L}}$ denotes the size and $u_{i,L}^T$ the reverse of the sequence $u_{i,L}$. Imagine now the full scenario depicted in fig.~\ref{fig:U1connection}: In a similar fashion to above we can ``flip'' the elements of $u_{i,L}$ and $v_{j,L}$ to the other sides of the quark lines $i$ and $j$, respectively. Connecting then the quark lines $i$ and $j$ is indeed $U(1)$--like, and we may thus define the operation $U_{ij}(u,v)$ equally well by
\begin{align}
U_{ij}(u,v)=(-1)^{n_{u_{i,L}}+n_{v_{j,L}}}\sum\limits_{w\in[q_i(u_{i,R} \shuffle u_{i,L}^T)\qbar_iq_j(v_{j,R} \shuffle v_{j,L}^T)\qbar_j]} w \;.
\end{align}

\section{Tree-level multi-quark amplitudes}

We concentrate on the case of amplitudes $\hat{\scriptA}$ with distinct quark pairs, since the case of amplitudes with identical quark pairs can always be related to the former by antisymmetrization \cite{Reuschle:2013qna,Weinzierl:2005dd}. The color decomposition of a tree--level amplitude with $n_g$ gluons and $n_q$ distinct quark pairs reads
\begin{align}
&{\hat\scriptA}^{(0)}_n = 
\left(\frac{g_s}{\sqrt{2}}\right)^{n-2}
\sum\limits_{\pi \in S_{n_q}}
\sum\limits_{\stackrel{i_1,...,i_{n_q} \ge 0}{i_1+...+i_{n_q}=n_g}}
\sum\limits_{\sigma \in S_{n_g}}
c_{\mathrm{open}}\big(q_1,g_{\sigma_1},...,g_{\sigma_{i_1}},\bar{q}_{\pi_1}\big)
c_{\mathrm{open}}\big(q_2,g_{\sigma_{i_1+1}},...,g_{\sigma_{i_1+i_2}},\bar{q}_{\pi_2}\big)
\nonumber \\[-0.75ex]
&...
c_{\mathrm{open}}\big(q_{n_q},g_{\sigma_{i_1+...+i_{n_q-1}+1}},...,g_{\sigma_{i_1+...+i_{n_q}}},\bar{q}_{\pi_{n_q}}\big)
A^{(0)}_{n}\big(q_1,g_{\sigma_1},...,g_{\sigma_{i_1}},\bar{q}_{\pi_1}, q_2, ..., g_{\sigma_{i_1+...+i_{n_q}}}, \bar{q}_{\pi_{n_q}} \big) \;,
\end{align}
which generates all possible distinct color--flow combinations by summing over all permutations $\pi$ of $n_q$ antiquark indices, all partitions $\{i_1,...,i_{n_q}\}$ of $n_g$ gluon indices among $n_q$ open strings, and all permutations $\sigma$ of $n_g$ gluon indices.
Without loss of generality we can relabel the quarks and antiquarks such that the permutation $\pi$ of antiquark indices can be written in terms of $r$ cycles with sizes $k_j$ ($j=1,...,r$):
\begin{align}
\pi = 
\left(1,2,...,k_1\right) 
\left(k_1+1,...,k_1+k_2\right)
\ldots
\left(k_1+...+k_{r-1}+1,...,k_1+...+k_r\right) \;,
\label{eq:picycles}
\end{align}
where $k_1+...+k_r=n_q$. Each cycle $j$ in $\pi$ corresponds directly to a product of open strings and defines a cyclic word
{\small\begin{align}
u_j = 
\left[ q_{k_1+...+k_{j-1}+1} ... \bar{q}_{k_1+...+k_{j-1}+2}
q_{k_1+...+k_{j-1}+2} ... \bar{q}_{k_1+...+k_{j-1}+3}
\ldots
q_{k_1+...+k_{j-1}+k_j} ... \bar{q}_{k_1+...+k_{j-1}+1} \right] \;,
\label{eq:picyclicword}
\end{align}}%
where every cyclic word simply contains an ordered sequence of quark, antiquark and gluon indices, corresponding to the members of a color--connected cluster. Each partial amplitude can thus be associated to a set $\{u_1,...,u_r\}$ of $r$ cyclic words, corresponding to the associated permutation $\pi$ of antiquark indices. Two cyclic words are thereby connected by a $U(1)$ gluon, and each partial amplitude contains therefore $(r-1)$ $U(1)$ gluons. 
We can then use the shuffle operation $U(u_1,...,u_r)$, defined in the previous section, 
to determine all cyclic orderings to this partial amplitude, such that its decomposition into primitive amplitudes reads
\begin{align}
A_n^{(0)}=\left(-\frac{1}{N}\right)^{r-1} \sum\limits_{w\in U(u_1,...,u_r)}P_n^{(0)}(w) \;.
\label{eq:treelevelmultiquarkpartialtoprimitive}
\end{align}

\section{Basic one-loop operations}

\begin{figure}
\vspace{-0.75cm}
\centering
\includegraphics[scale=0.625]{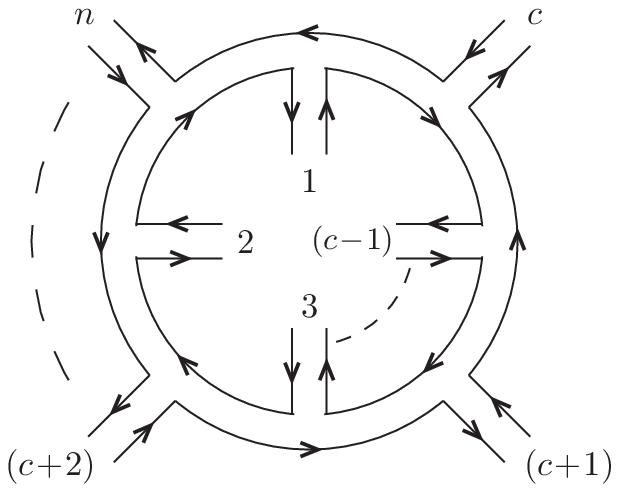}
\scalebox{0.625}{
\begin{picture}(180,180)(-90,-90)
%
%
\SetArrowInset{0.0}
\SetColor{Red}
\SetWidth{2.5}
\Arc[flip,arrow,arrowlength=8,arrowwidth=4,arrowpos=0.75](0,0)(52.5,135,225)
\Line[arrow,arrowlength=7,arrowwidth=3.5,arrowpos=0.7](-35.0,37.5)(-58.445,61.445)
\Line[flip,arrow,arrowlength=7,arrowwidth=3.5,arrowpos=0.7](-35.0,-37.5)(-58.445,-61.445)
\SetArrowInset{1.0}
\SetColor{Black}
\SetWidth{0.5}
%
%
\Arc[flip,arrow,arrowlength=5,arrowwidth=2](0,0)(50,5,85)
\Line[flip,arrow,arrowlength=5,arrowwidth=2](5,50)(5,25)
\Line[flip,arrow,arrowlength=5,arrowwidth=2](-5,25)(-5,50)
\Text(0,17.5)[]{$3$}
\Arc[flip,arrow,arrowlength=5,arrowwidth=2,arrowpos=0.25](0,0)(50,95,175)
\Line[flip,arrow,arrowlength=5,arrowwidth=2](-50,5)(-25,5)
\Line[flip,arrow,arrowlength=5,arrowwidth=2](-25,-5)(-50,-5)
\Text(-18.5,0)[]{$4$}
\Arc[flip,arrow,arrowlength=5,arrowwidth=2,arrowpos=0.75](0,0)(50,185,265)
\Line[flip,arrow,arrowlength=5,arrowwidth=2](-5,-50)(-5,-25)
\Line[flip,arrow,arrowlength=5,arrowwidth=2](5,-25)(5,-50)
\Text(0,-18.5)[]{$5$}
\Arc[flip,arrow,arrowlength=5,arrowwidth=2](0,0)(50,275,355)
\Line[flip,arrow,arrowlength=5,arrowwidth=2](50,-5)(25,-5)
\Line[flip,arrow,arrowlength=5,arrowwidth=2](25,5)(50,5)
\Text(12.5,0)[]{\small$(c\!+\!1)$}
\Arc[dash,dashsize=5](0,0)(30,285,345)
%
%
\Arc[arrow,arrowlength=5,arrowwidth=2](0,0)(60,320,40)
\Line[flip,arrow,arrowlength=5,arrowwidth=2](38.57,45.96)(56.25,63.64)
\Line[arrow,arrowlength=5,arrowwidth=2](45.96,38.57)(63.64,56.25)
\Text(75,65)[]{$(c\!+\!2)$}
\Arc[arrow,arrowlength=5,arrowwidth=2](0,0)(60,50,130)
\Line[arrow,arrowlength=5,arrowwidth=2,arrowpos=0.3](-38.57,45.96)(-56.25,63.64)
\Text(-65,65)[]{$2$}
\Line[flip,arrow,arrowlength=5,arrowwidth=2,arrowpos=0.3](-38.57,-45.96)(-56.25,-63.64)
\Text(-65,-65)[]{$1$}
\Arc[arrow,arrowlength=5,arrowwidth=2](0,0)(60,230,310)
\Line[arrow,arrowlength=5,arrowwidth=2](38.57,-45.96)(56.25,-63.64)
\Line[flip,arrow,arrowlength=5,arrowwidth=2](45.96,-38.57)(63.64,-56.25)
\Text(65,-65)[]{$n$}
\Arc[dash,dashsize=10](0,0)(90,325,35)
\end{picture}
}
\hspace{0.9cm}
\rule[0.5cm]{0.5pt}{3.0cm}
\hspace{0.65cm}
\raisebox{0.2cm}{\includegraphics[scale=1.025]{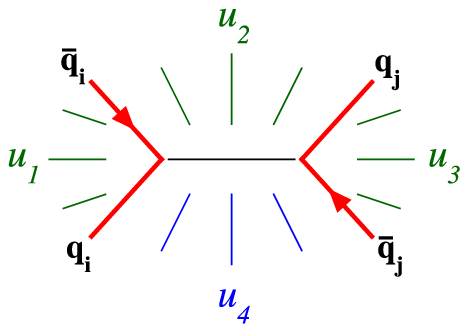}}
\vspace{-0.5cm}
\caption{\footnotesize The two diagrams on the left depict examples for color--flow diagrams corresponding to $c_{\mathrm{closed}}\big(g_{1},...,g_{c-1}\big)c_{\mathrm{closed}}\big(g_{c},...,g_{n}\big)$ and $c_{\mathrm{closed}}\big(g_{3},...,g_{c+1}\big)c_{\mathrm{open}}\big(q_2,g_{c+2},...,g_{n},\qbar_1\big)$, respectively. The diagram in the middle corresponds thereby for example to a contribution with one quark pair, where the associated continuous fermion line is indicated in red. The diagram on the very right promotes two quark lines $i$ and $j$ in the cyclic word $u=[q_iu_1\qbar_iu_2q_ju_3\qbar_ju_4]$, where the sequences $u_1$ through $u_4$ may contain further quark and antiquark indices, as well as gluon indices. The bold lines in red correspond to the continuous fermion lines of the two quark pairs.}
\label{fig:U1inloop}
\end{figure}

In one--loop amplitudes we encounter color flows of double--ring structure, with emissions from an inner and an outer ring. Consider these structures to correspond to two cyclic words $u=[l_1...l_k]$ and $v=[l_{k+1}...l_r]$, where emission from the outer ring is in clockwise order of $u$, while emission from the inner ring is in anti--clockwise order of $v$, as for example depicted in the left diagram of fig.~\ref{fig:U1inloop} with $u=[g_{c}...g_{n}]$ and 
$v=[g_{1}...g_{c-1}]$. 
If we want to find all cyclic orderings corresponding to the color flow of the associated contribution, we may use the cyclic shuffle product:
\begin{align}
(-1)^{r-k}u\circledcirc v^T \;.
\end{align}

Further we can have exchange of $U(1)$ gluons between tree--like quark lines and loop structures, where we distinguish two cases:
\vspace{-1.0ex}
\begin{enumerate}
\itemsep-0.5ex
\item[(i)] A quark line color--connected to the loop, in which case there is a cyclic word $u_j^{[l]}$, with $l=1\!/2,1$, color--connected to the loop (regarding the label $[l]$, see sec.~\ref{sec:oneloopmultiquark}). In this case the tree--level operation $U$ on the cyclic words $\{u_{i\neq j}\}$ and $u_j^{[l]}$ suffices: $U( u_1, ..., u_{j-1}, u_j^{[l]}, u_{j+1}, ..., u_{r} )$. 
\item[(ii)] A closed fermion loop, in which case there is a cyclic word $v^{[1\!/2]}$, corresponding to a separate color cluster. 
In this case the tree--level operation $U$ on the cyclic words $\{u_{i}\}$ and $v^{[1\!/2]}$ suffices: $U\left(u_1, ..., u_{r}, v \right)$.
\end{enumerate}

Finally we can have $U(1)$ gluons in the loop, where we distinguish three cases:
\vspace{-1.0ex}
\begin{enumerate}
\itemsep-0.5ex
\item[(i)] A $U(1)$ gluon closing on the same quark line: Consider $u=[q_iu_{i,R}\qbar_iu_{i,L}]$, as depicted by the left cyclic word in fig.~\ref{fig:U1connection}. The loop shall be closed by a $U(1)$--like connection, which is implemented by the operation
\begin{align}
C_{ii}(u)=(-1)^{n_{u_{i,L}}}\sum\limits_{w\in[q_i(u_{i,R}\shuffle u_{i,L}^T)\qbar_i]}w \;,
\end{align}
where we essentially ``flip'' the elements of $u_{i,L}$ to the other side of the quark line $i$. Closing the loop on the ``lower'' side is then $U(1)$--like. The definition of $C_{ij}(u)$ is not unique.
\item[(ii)] A $U(1)$ gluon closing between two quark lines of the same cyclic word $u\!=\![q_iu_{\!1}\!\qbar_iu_2q_{\!j}u_3\qbar_{\!j}u_4]$, as depicted on the very right of fig.~\ref{fig:U1inloop}. The loop shall be closed by a $U(1)$--like connection between the quark lines $i$ and $j$, which is implemented by the operation
\begin{align}
C_{ij}^{RR}(u)=(-1)^{n_{u_{4}}}\sum\limits_{w\in[q_i^R((u_{1}\qbar_i^Ru_2q_j^Ru_3)\shuffle u_{4}^T)\qbar_j^R]}w \;,
\label{eq:loopclosingRR}
\end{align}
where we essentially ``flip'' the elements of $u_{4}$ to the other side of the $q_i$--$q_j$ line. Closing the loop on the ``lower'' side between the quark lines $i$ and $j$ is then $U(1)$--like. There are in total four categories of this type, corresponding to the different ``routing'' combinations of the fermion--flow arrows of $i$ and $j$ with respect to the loop, which we need to sum in the end: $C_{ij}(u)=C_{ij}^{RR}(u)+C_{ij}^{RL}(u)+C_{ij}^{LR}(u)+C_{ij}^{LL}(u)$. Again, their definition is not unique.
\item[(iii)] Multiple $U(1)$ gluons closing between quark lines of different cyclic words: Cutting one $U(1)$ gluon gives a tree and we expect that the remaining $U(1)$ gluons can be treated with the operation $U_{ij}$. 
We define an operation $CU(u_1,...,u_r)$, acting on $r$ cyclic words, which combines the operation $U_{ij}$ and $C_{ij}$ by the following algorithm:
\vspace{-1.0ex}
\begin{enumerate}
\itemsep-0.5ex
\item[1.] Draw all connected one--loop diagrams with $r$ vertices, labeled by $u_1$, ..., $u_r$, including diagrams with self--loops. Select in each diagram one edge $e_C$, such that upon removal of this edge the diagram becomes a connected tree diagram.
\item[2.] For every edge $e \neq e_C$ connecting $u_k$ and $u_l$ use an operation $U_{ij}$, where $i$ labels a quark pair in $u_k$ and
$j$ labels a quark pair in $u_l$ and sum over $i$ and $j$.
The order in which these operations are performed is irrelevant, since the operations for a given graph are associative.
\item[3.] If $e_C$ is not a self--loop and connects $u_k$ and $u_l$ use an operation $C_{ij}$, where $i$ labels a quark pair in $u_k$ and
$j$ labels a quark pair in $u_l$ and sum over $i$ and $j$.
\item[4.] If $e_C$ is a self--loop connected to $u_k$ use an operation $C_{ij}$, where $i$ and $j$ label quark pairs in $u_k$ and sum over $i$ and $j$ subject to $i \le j$.
\item[5.] Divide by the symmetry factor of the diagram. Sum over all diagrams.
\end{enumerate}
\end{enumerate}

\section{One-loop multi-quark amplitudes}
\label{sec:oneloopmultiquark}

The color decomposition of a one--loop amplitude with $n_g$ gluons and $n_q$ distinct quark pairs reads
\begin{align}
&{\hat\scriptA}^{(1)}_n = 
\left(\frac{g_s}{\sqrt{2}}\right)^{n}
\sum\limits_{\pi \in S_{n_q}}\!\!
\sum\limits_{\stackrel{i_1,...,i_{n_q},m \ge 0}{i_1+...+i_{n_q}+m=n_g}}\!\!
\sum\limits_{\sigma \in S_{n_g}\!/Z_m}\!\!
c_{\mathrm{open}}\big(q_1,g_{\sigma_1},...,g_{\sigma_{i_1}},\bar{q}_{\pi_1}\big)
c_{\mathrm{open}}\big(q_2,g_{\sigma_{i_1+1}},...,g_{\sigma_{i_1+i_2}},\bar{q}_{\pi_2}\big)
\nonumber \\[-0.75ex]
&\hspace{4.0cm}\ldots\,
c_{\mathrm{open}}\big(q_{n_q},g_{\sigma_{i_1+...+i_{n_q-1}+1}},...,g_{\sigma_{i_1+...+i_{n_q}}},\bar{q}_{\pi_{n_q}}\big)
c_{\mathrm{closed}}\big(g_{\sigma_{i_1+...+i_{n_q}+1}},...,g_{\sigma_{n_g}}\big)
\nonumber\\[0.75ex]
&\hspace{4.0cm}
A^{(1)}_{n,m}\big(q_1,g_{\sigma_1},...,g_{\sigma_{i_1}},\bar{q}_{\pi_1}, q_2, ..., g_{\sigma_{i_1+...+i_{n_q}}}, \bar{q}_{\pi_{n_q}}; g_{\sigma_{n_g-m+1}},...,g_{\sigma_{n_g}} \big) \;,
\end{align}
which generates all possible distinct color--flow combinations by summing over all permutations $\pi$ of $n_q$ antiquark indices, all partitions $\{i_1,...,i_{n_q},m\}$ of $n_g$ gluon indices among $n_q$ open strings and one closed string, and all permutations $\sigma$ of $n_g$ gluon indices, except those that leave the closed string invariant.
Again, the permutation $\pi$ of antiquark indices can be written in terms of $r$ cycles, as given in eq.~(\ref{eq:picycles}), where each cycle defines a cyclic word, as given in eq.~(\ref{eq:picyclicword}). 
The additional closed string defines an additional cyclic word $v$, which contains $m$ gluon indices:
\begin{align}
c_{\mathrm{closed}}\big(g_{\sigma_{n_g-m+1}},...,g_{\sigma_{n_g}}\big)
\Rightarrow
v = \big[g_{\sigma_{n_g-m+1}} ... g_{\sigma_{n_g}}\big] \;.
\end{align}

One--loop partial and primitive amplitudes, as already mentioned, are further characterized by two additional properties: 
{\it1st)} Additional routing labels $L$ or $R$ distinguish whether fermion lines (following their fermion--flow arrows) turn left or right with respect to the loop. The reflection identity states then that $P^{(1)}(w)=(-1)^nP^{(1)}(w^T)$, while $\{L\leftrightarrow R\}$. In the following we will not concentrate any further on the assignment of routing labels. 
{\it2nd)} For each partial amplitude we separate contributions with a closed fermion loop, which we label by ${[1\!/2]}$, from those without, which we label by ${[1]}$: $A^{(1)} = A^{(1)[1\!/2]} + A^{(1)[1]}$. Primitive amplitudes are to be labeled accordingly.

The decomposition of the partial amplitudes $A^{(1)[1\!/2]}$ into primitive amplitudes reads
\begin{align}
A^{(1)[1\!/2]}_{n,m} 
&= \frac{\delta_{m,0}}{N} N_f
\left(-\frac{1}{N}\right)^{r-1}
\sum\limits_{j=1}^{r}
\sum\limits_{w \in U\big( u_1, ..., u_{j-1}, u_j^{[1\!/2]}, u_{j+1}, ...,  u_{r}\big)} 
P^{(1) [1\!/2]}_n\left( w \right)
\nonumber\\[1.0ex]
&+ \hspace{0.8cm}N_f
\left(-\frac{1}{N}\right)^{r}
\sum\limits_{w \in U\left(u_1, ..., u_{r}, v \right)} 
P^{(1) [1\!/2]}_n\left( w \right) \;,
\label{nf_decomposition}
\end{align}
where in general we may have $N_f$ flavors running in the loop and $c_{closed}(...)$ may be empty. If $\pi$ consists of $r$ cycles, we can have either $(r-1)$ or $r$ tree--like $U(1)$ gluons in the amplitude.
The first term in eq.~(\ref{nf_decomposition}) corresponds to contributions where one of the color clusters, which we denote by $u_j^{[1\!/2]}$, is color--connected to the closed fermion loop, i.e. it contributes to the case $m=0$. We need, however, an explicit factor $\frac{1}{N}$ to compensate for the factor $c_{closed}()=N$, which is not present in these contributions. There are $(r-1)$ tree--like $U(1)$ gluons, which connect between $\{u_{i\neq j}\}$ and $u_j^{[1\!/2]}$. Since each of the color clusters can be the one color--connected to the closed fermion loop, we need to sum over $j$. 
The second term in eq.~(\ref{nf_decomposition}) corresponds to contributions where none of the external quark lines is color--connected to the closed fermion loop. The closed fermion loop corresponds to a closed string with gluons attached and thus belongs to a separate color cluster, associated to the cyclic word $v$, i.e. it contributes to the case $m\geq0$. 
We need no explicit factor $\frac{1}{N}$ here, since any factor $c_{closed}()=N$ generated for $m=0$, is genuine to these contributions. There are $r$ tree--like $U(1)$ gluons, which connect between $\{u_i\}$ and $v$. In every $P^{(1)[1\!/2]}_n\left( w \right)$ an additional minus sign needs to be included, due to the closed fermion loop.

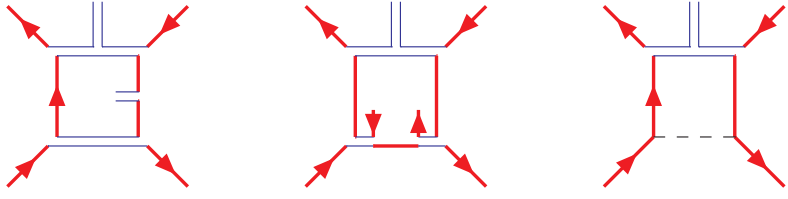
\begin{figure}
\vspace{-0.6cm}
\centering
\scalebox{0.85}{
\begin{picture}(130,100)(0,0)
\SetArrowInset{0.0}
\SetColor{Red}
\SetWidth{1.5}
\ArrowLine(28,72)(10,90)
\ArrowLine(10,10)(28,28)
\ArrowLine(72,28)(90,10)
\ArrowLine(90,90)(72,72)
\ArrowLine(32,32)(32,68)
\SetColor{Blue}
\SetWidth{0.5}
\Line(32,68)(68,68)
\SetColor{Red}
\SetWidth{1.5}
\Line(68,68)(68,52)
\Line(68,48)(68,32)
\SetColor{Blue}
\SetWidth{0.5}
\Line(68,32)(32,32)
\Line(72,72)(52,72)
\Line(48,72)(28,72)
\Line(28,28)(72,28)
\Line(48,72)(48,92)
\Line(52,72)(52,92)
\Line(68,48)(58,48)
\Line(68,52)(58,52)
\end{picture}
\begin{picture}(130,100)(0,0)
\SetArrowInset{0.0}
\SetColor{Red}
\SetWidth{1.5}
\ArrowLine(28,72)(10,90)
\ArrowLine(10,10)(28,28)
\ArrowLine(72,28)(90,10)
\ArrowLine(90,90)(72,72)
\SetColor{Blue}
\SetWidth{0.5}
\SetColor{Red}
\SetWidth{1.5}
\Line(32,32)(32,68)
\SetColor{Blue}
\SetWidth{0.5}
\Line(32,68)(68,68)
\SetColor{Red}
\SetWidth{1.5}
\Line(68,68)(68,32)
\SetColor{Blue}
\SetWidth{0.5}
\Line(68,32)(60,32)
\Line(40,32)(32,32)
\SetColor{Red}
\SetWidth{1.5}
\ArrowLine(40,44)(40,32)
\ArrowLine(60,32)(60,44)
\SetColor{Blue}
\SetWidth{0.5}
\Line(72,72)(52,72)
\Line(48,72)(28,72)
\Line(28,28)(40,28)
\SetColor{Red}
\SetWidth{1.5}
\Line(40,28)(60,28)
\SetColor{Blue}
\SetWidth{0.5}
\Line(60,28)(72,28)
\Line(48,72)(48,92)
\Line(52,72)(52,92)
\end{picture}
\begin{picture}(130,100)(0,0)
\SetArrowInset{0.0}
\SetColor{Red}
\SetWidth{1.5}
\ArrowLine(28,72)(10,90)
\ArrowLine(10,10)(32,32)
\ArrowLine(68,32)(90,10)
\ArrowLine(90,90)(72,72)
\ArrowLine(32,32)(32,68)
\SetColor{Blue}
\SetWidth{0.5}
\Line(32,68)(68,68)
\SetColor{Red}
\SetWidth{1.5}
\Line(68,68)(68,32)
\SetColor{Black}
\SetWidth{0.5}
\DashLine(68,32)(32,32){5}
\SetColor{Blue}
\Line(72,72)(52,72)
\Line(48,72)(28,72)
\Line(48,72)(48,92)
\Line(52,72)(52,92)
\end{picture}
}
\vspace{-0.35cm}
\caption{\footnotesize Examples of color--flow diagrams for the three cases discussed in eq.~(\protect\ref{decomposition}). The first diagram shows an example with a closed string on the inner ring (case 1). The second diagram shows an example without a closed string. Here, particles attached to the outer ring are colour--disconnected from particles attached to the inner ring (case 2). The third diagram shows also an example without a closed string but with a $U(1)$ gluon in the loop. Here, particles attached to the outer ring are colour--connected to particles attached to the inner ring (case 3). The bold lines in red indicate the continuous fermion lines of quark pairs, the double lines correspond to $U(N)$ gluons, the dashed lines to $U(1)$ gluons.}
\label{figure_one_loop_1}
\end{figure}

The decomposition of the partial amplitudes $A^{(1)[1]}$ into primitive amplitudes reads
\begin{align}
A^{(1)[1]}_{n,m}
&=
 \left(-1\right)^m
 \left(-\frac{1}{N}\right)^{r-1}
 \sum\limits_{j=1}^r
 \sum\limits_{w \in U\left( u_1, ..., u_{j-1}, u_j^{[1]}, u_{j+1}, ..., u_{r} \right) \circledcirc v^T} 
 P^{(1) [1]}_{n}\left(w\right)
 \nonumber\\[1.0ex]
&+
 \frac{\delta_{m,0}}{N}\hspace{0.4cm}
 \left(-\frac{1}{N}\right)^{r-2}
 \sum\limits_{i=1}^{r-1}
 \sum\limits_{j=i+1}^r
 \left(-1\right)^{n_{u_j}}
 \sum\limits_{w \in U\big( u_1, ..., u_i\!\!\!\!\!\!\!\diagdown, ..., u_j\!\!\!\!\!\!\!\diagdown, ..., u_{r}, \big( u_i \circledcirc u_j^T \big)^{\!\![1]\;} \big)} 
 P^{(1) [1]}_{n}\left(w\right)
 \nonumber\\[1.0ex]
&+
 \frac{\delta_{m,0}}{N}\hspace{0.4cm}
 \left(-\frac{1}{N}\right)^{r}
 \!\!\!\!\!
 \sum\limits_{w \in CU\left(u_1, ..., u_{r}\right)} 
 P^{(1) [1]}_{n}\left(w\right)
\label{decomposition}
\end{align}
where we have to deal with double--ring structures now, and where $u_k\!\!\!\!\!\!\!\diagdown$ denotes that the cyclic word $u_k$ is to be taken off the list of arguments of $U(...)$. If $\pi$ consists of $r$ cycles, we can have either $(r-1)$, $(r-2)$ or $r$ $U(1)$ gluons in the amplitude.
The first term in eq.~(\ref{decomposition}) corresponds to contributions where we have a closed string with $\geq0$ gluons attached, i.e. it contributes to the case $m\geq0$ (we define the closed string to be on the inner ring). One of the cyclic words $u$ corresponds to the loop, which we denote by $u_j^{[1]}$, while the others correspond to tree--like structures. There are $(r-1)$ tree--like $U(1)$ gluons, which connect between the $r$ cycles. Since each of the color clusters can be the one color--connected to the loop, we need to sum over $j$. 
The second term in eq.~(\ref{decomposition}) corresponds to contributions where we have no closed string, i.e. it contributes to the case $m=0$. Two of the cyclic words $u$ correspond now to the loop, where one is associated to the inner ring and the other to the outer ring, and where particles on the inner ring are color--disconnected from particles on the outer ring. There are $(r-2)$ tree--like $U(1)$ gluons, which connect between $\{u_{k\neq i,j}\}$ and $\big(\! u_i \!\circledcirc\! u_j^T \!\big)^{\!\![1]}$, and we need to sum over $i$ and $j<i$.
The third term in eq.~(\ref{decomposition}) corresponds also to contributions where we have no closed string, i.e. it contributes also to the case $m=0$. There are $r$ $U(1)$ gluons, of which at least one is in the loop, and particles on the inner ring are color--connected to particles on the outer ring. Here we use the operation $CU(u_1,...,u_r)$, defined in the previous section. Examples of color--flow diagrams in all three cases are shown in fig.~\ref{figure_one_loop_1}.

\section{Summary}

The color--flow decomposition yields a systematic approach to complicated color structures. The reduction to primitive amplitudes, particularly in the case of one--loop amplitudes with arbitary numbers of gluons and quarks, is thereby a non--trivial problem. Closed expressions can nevertheless be formulated through (generalized) shuffle relations.
We have validated the formulae that result thereby from our method to converge to the known results in \cite{oneloopcolordecompsimple}, for the case of one--loop amplitudes with one quark pair and an arbitrary number of gluons, and further compared to the results of Ita et al. in \cite{oneloopmultiquarkpragmatic}, e.g. for the case of one--loop amplitudes with two quark pairs plus two gluons. The corresponding checks and examples, 
as well as further comments as to the assignment of routing labels or the loop closing operation in eq.~(\ref{eq:loopclosingRR}) etc., 
can be found in \cite{Reuschle:2013qna}.
We want to stress that shuffle operations yield in general a decomposition into primitive amplitudes with a high degree of symmetry, but not necessarily with a minimum number of terms. However, by using identities between primitive amplitudes, e.g. the reflection identity, one can always reduce the number of terms.
Finally we want to note that the method, which we present here, offers the possibility to study subleading effects in the $1/N$ expansion of one--loop QCD amplitudes with an arbitrary number of quarks and gluons in a systematic way. Moreover we would like to suggest that shuffle operations might also be of help in order to study the color structure of two--loop amplitudes.


\end{document}